# DETERMINATION OF THE ENERGY CHARACTERISTICS OF AN ELECTRON BEAM USING A LIGHT SCINTILLATOR

V.I. Alekseev[a], V.A. Baskov[a]*, V.A. Dronov[a], A.I. L'vov[a], A.V. Koltsov[a], Yu.F. Krechetov[b], V.V. Polyansky[a]

*a - P.N. Lebedev Physical Institute, Moscow, Leninsky Prospekt, 53, 119991 Russia*
*b - Joint Institute for Nuclear Research, Dubna, Moscow Region, 6 Joliot-Curie street, 141980 Russia*
*\*E-mail: baskov@x4u.lebedev.ru*

The possibility of using the effect of full energy absorption in a light scintillator when an electron beam passes through it to determine the energy characteristics of a low and medium energy beam (the "absorbed energy" method) is experimentally presented. The energy calibration of the quasi-monochromatic electron beam of the Pakhra accelerator of the P. N. Lebedev Physical Institute of the Russian Academy of Sciences using scintillation detectors with a thickness of 14.5, 20, 23.5 and 51.2 cm was performed. For electron beam energies up to ~100 MeV and scintillation detector thicknesses from 5 to 20 cm the accuracy of electron beam energy determination was 10 to 20%, respectively.

## Introduction

The energy characteristics of the electron beam, including the maximum and average energy, as well as the energy spectrum, are the most important characteristics. Synchrotron, Cherenkov and transition radiation carry information about the energy parameters of the beam. The total energy of the beam is directly measured by the calorimetric method. Energy characteristics can be determined as one of the methods, as well several at the same time. Determination of the energy characteristics of the beam can be carried out with and without destroying the beam. For example, the method of laser scattering or the method of application of thin scintillation, Cherenkov and solid detectors does not destroy the beam, and the calorimetric method, for example, the use of the Cherenkov total absorption spectrometer, leads to the destruction of the beam [1].

This paper shows the possibility of determining the energy characteristics of the electron beam using the method of "absorbed energy". The principle of the method is to determine the energy characteristics of the electron beam by changing the thickness L of scintillation detectors or changing the energy $E_e$ of the electron beam to a dimension at which the trajectory of individual electrons completely fits into the volume and thickness of the detector. In this case, the electron energy of the beam corresponds to the sum of the average ionization losses of an electron per unit path in the detector $<E> = k \cdot L$, where $k = \Delta E/\Delta x$ ($\Delta E/\Delta x$ is the average ionization losses of electrons per unit path in the detector; L - thickness of the scintillation detector) [2].

**Experimental setup**

The research was carried out on a quasi-monochromatic beam of secondary electrons of the accelerator "Pakhra" of the Lebedev Physical Institute of the Russian Academy of Sciences (LPI) (Fig. 1). The converter was a copper plate with thickness of 3 mm and 3.2 mm in diameter, located on the "slice" of the magnet poles [3]. The trigger signal T was the signal from the coincidence of the polystyrene scintillation counters $S_1 - S_3$ signals and the veto counter A with an orifice of ⌀10 mm (T=( $S_1 \cdot S_2 \cdot S_3$)·A). The counters $S_1 - S_3$ and A were size of 15×15×1 mm³ and 60×90×10 mm³, respectively. The intensity of the secondary electron beam was ~$10^2$ e⁻/sec.

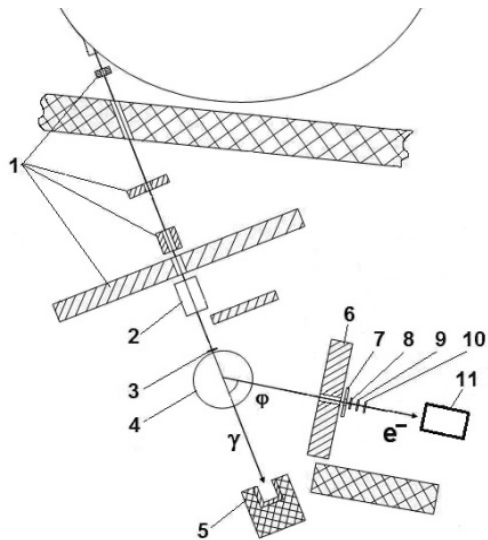

**Fig. 1.** The diagram of the calibration quasi-monochromatic secondary-electron beam: 1 - lead collimators; 2 - cleaning magnet SP-3; 3 - converter; 4 - magnet SP-57; 5 - photon-beam absorber ("burial ground"); 6 - collimator (ø10 mm); 7 - scintillation counter "veto" A; 8 - 9 - trigger scintillation counters $S_1$ - $S_3$; 11 - scintillation detector (SD).

Two scintillation detectors (SD) made of polystyrene of 20×20×20 cm$^3$ (SD1) and 14.5×23.5×51.2 cm$^3$ (SD2) were used (Fig. 2). Scintillators were examined by an assemblage of 7 photoelectron multipliers (PMT) of the PMT-85 type. PMT assemblages without lubricant were tightly pressed to the surface of the scintillators. The scintillators, excluding the area occupied by the PMT photocathode, were wrapped in metallic Mylar and black paper. The increase in multichannel on the basis of PMT-85 is associated with the preservation of the scintillation detector`s response in comparison, for example, with PMT-49 and not impairment of energy resolution. In the future, it is planned to use SD in the installation to register the effect in the presence of a significant low-energy electromagnetic background of ~ $10^4$-$10^5$ particles/sec. Amplitude SD was the sum of signals from all PMT assemblies except the constant component of the charge-to-digital converter (QDC) ("pedestal") of each channel. It can be assumed that the insignificant value of the secondary electron current (I ~ 0.03 nA [4]) practically does not affect on the value of the SD signal amplitude but no special study was conducted.

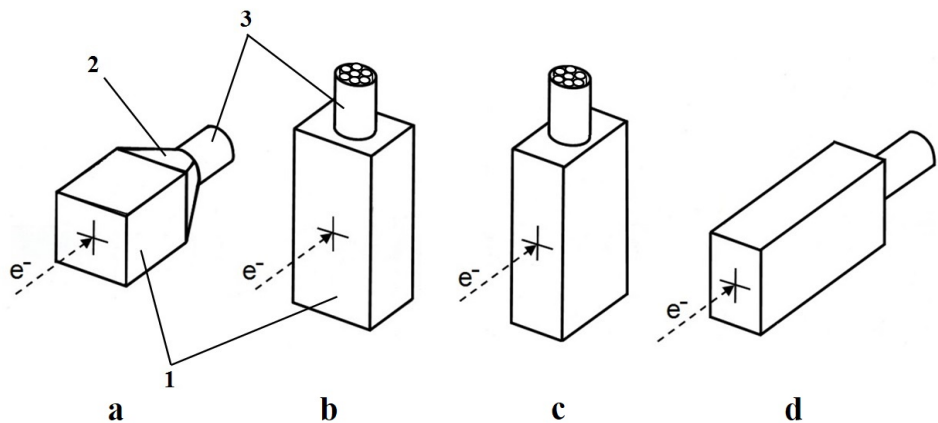

**Fig. 2.** The schematic view of scintillation detectors: a) – scintillation detector (SD1) of 20×20×20 cm$^3$ (beam thickness 20 cm); b)-scintillation detector (SD2) of 14.5×23.5×51.2 cm$^3$ (beam thickness 14.5 cm); C) – scintillation detector (SD2) of 14.5×23.5×51.2 cm$^3$ (beam thickness 23.5 cm); d) – scintillation detector (SD2) of 14.5×23.5×51.2 cm$^3$ (beam thickness 51.2 cm) (1 – scintillation unit; 2-light collector; 3 – assembly of 7 photoelectron multipliers PMT – 85).

The scheme of research is presented in Fig. 2. At the first phase (a) the energy characteristics of SD1 were researched, when the thickness of the counter beam were 20 cm. At the second phase (b), the beam had SD2 with a thickness of 14.5 cm. At the third phase (c) was SD2 with the beam`s thickness of 23.5 cm and the characteristics of SD2 with the beam`s thickness of 51.2 cm were tested at the last fourth phase (d).

Electronic flow chart of measurements is presented in Fig. 3. Signals with $S_1$ – $S_3$ duration t =10 ns were fed to constant-fraction discriminators (CFD) $CFD_1$ – $CFD_3$ (threshold voltage of the CFD was $U_{th1} = U_{th2} = U_{th3} = 30$ mV) and then through delay units $DU_1$– $DU_3$ were fed to the CC coincidence circuit. The input "Anti" was routed a signal from the counter "Veto" A duration of 100 ns, formed by $CFD_4$. The signal with CC was a trigger signal T ≡ "Start" to activation the assembly unit 8 of the input charge-to-digital converter (QDC), with the help of which the signals from the scintillation detector were "recorded" through the crate-controller of the CAMAC system.

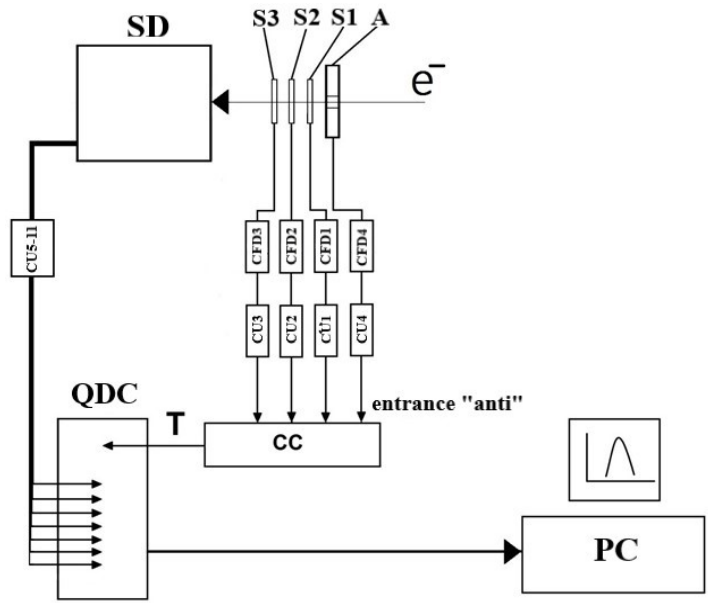

**Fig. 3.** The structured block diagram of measurements of the scintillation detector energy characteristics: $DU_1$-$DU_{11}$ - delay unit; $CFD_1$-$CFD_4$ - constant-fraction discriminator; CC - 4-channel coincidence circuit; QDC – charge-to-digital converter; PC-personal computer; SD - scintillation detector.

**Results**

The typical dependence of the amplitude SD1 with thickness of 20 cm on the electron energy is presented in Fig. 4. The figure demonstrates that at an electron energy of 40 MeV there is an abrupt change in the dependence. In the future, when the electron energy increases, the magnitude of the registered energy will change weakly. This means that at a detector thickness of 20 cm, the average ionization losses of electrons were $<E_e> = (\Delta E/\Delta x)\cdot L = 2$ [MeV/cm]·20[cm] = 40 MeV and do not increase with increase of energy (for the SD used in this work, the ionization losses were $\Delta E/\Delta x \approx 2$ MeV/cm [2]).

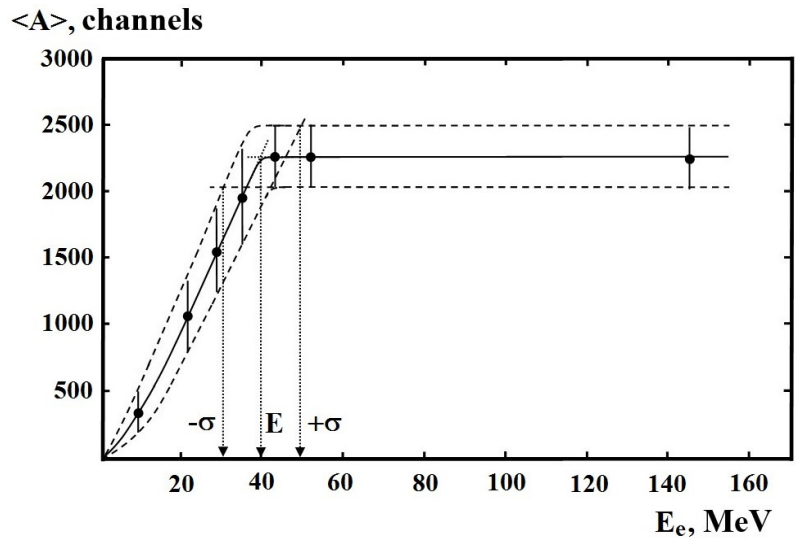

**Fig. 4.** Dependence of the average amplitude of the scintillation detector signal SD1 (<A>) from the energy of the quasi-monochromatic electron beam ($E_e$) (E - energy of the electron beam, determined by the method of " absorbed energy", σ-the error of the beam energy).

The dependence of the SD amplitude for all measured thicknesses on the energy of the secondary electron beam is presented in Fig. 5. Evidently that the dependence 1 determines the thickness of the SD when the electron tracks are inside the volume SC, and the dependence 2-5 determine the situation when the electron tracks of the beam go beyond the SD. Therefore, the dependence 1 can be called the "absorbed energy curve", and the point of sharp change in the dependence "inflection point".

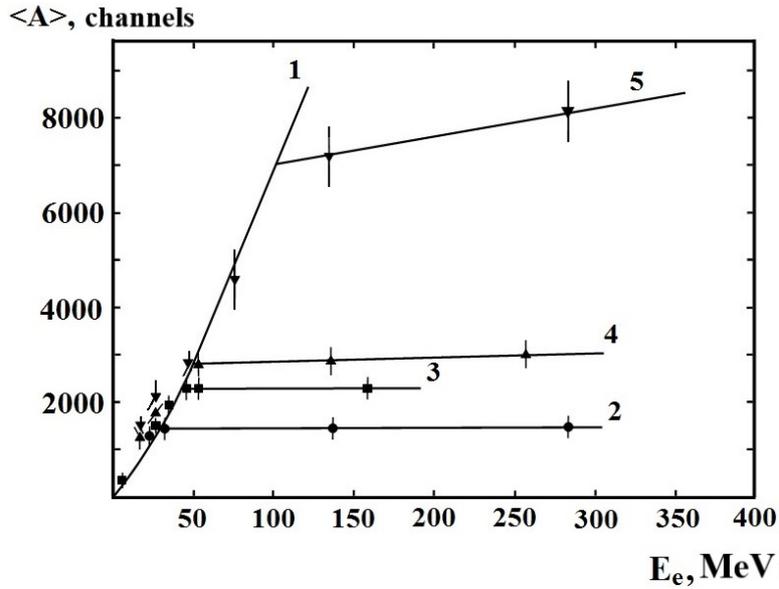

**Fig. 5.** Dependence of the average signal amplitudes of scintillation detectors SD1 and SD2 (<A>) from the energy of the quasi–monochromatic electron beam ($E_e$): 1 – the energy of the electron beam to the "saturation point"; 2 - 5-the energy of the electron beam after the "saturation point" (thickness SD1 and SD2 beam is 14.5, 20, 23.5 and 51.2 cm, respectively).

The Fig. 6 illustrates the dependence of the amplitude of the SD signal on the thickness of the scintillator at the "inflection point" (Fig. 5). Clearly that within the investigated thickness of the SD the dependence has a linear character. However, when extrapolation of the dependence into the energy region is close to zero, the dependence ceases to be linear.

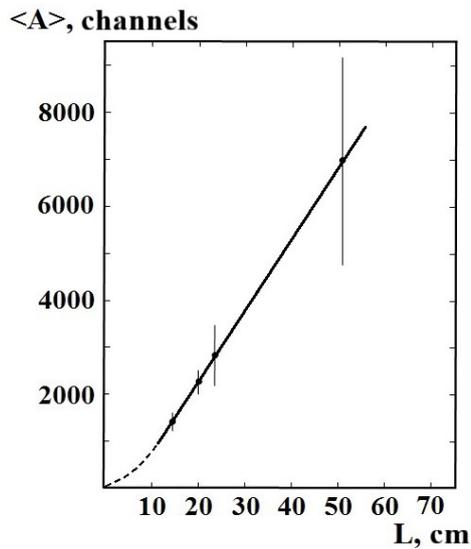

**Fig.6.** Dependence of average amplitudes of signals from scintillation detectors SD1 and SD2 (<A>) from the thickness of the detectors in the beam (L).

The Fig. 7 demonstrates the final dependence of the electron beam energy determined by this method on the electron energy determined by the estimate of the average ionization losses at the corresponding SD thickness. It is seen that the dependence is linear and within the error limits (in this case, the energy resolution errors include the influence of the copper Converter forming the electron beam [3] and the energy resolution of the SD as a detector). The beam energy values determined experimentally coincide with the energy values determined by the estimates. In addition, the values of the beam energy determined experimentally by this method coincide with the values of the energy of the secondary electron beam formed on the basis of the brake photon beam by the magnetic system and detected by the SD (figures 1 and 4) [3].

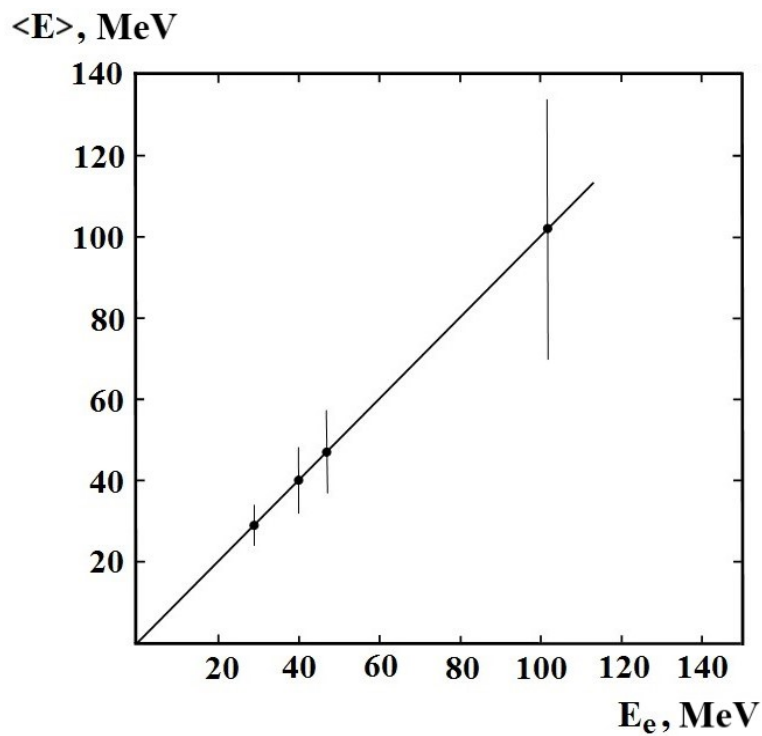

**Fig.7** Dependence of the electron beam energy (E), determined by the method of "absorbed energy", from the electron beam energy (E).

The electron energy error at this point was determined by the extrapolation of the energy errors determined before and after "the inflection point" to "the inflection point" along the corresponding trajectories "E – σ" and "E + σ" (Fig. 4).

For a detector thickness of 20 cm, the beam energy at the "the inflection point" was E = 40 ± 10 MeV.

Preliminary calibration of both SD fulfilled on isolated cosmic muons by the "through" method [3] demonstrated that the energy resolution of SD with a thickness of 20 cm is $\sigma \approx 9\%$ ($\sigma = \Delta E/E/2.35$, $\Delta E$ - the full width of the energy spectrum of cosmic muons at half its height, E - the average energy release of cosmic muons in the SD).

Thus, except the energy resolution of the preliminary calibration, which determines the energy resolution of the SD as a device, the energy value of the secondary electron beam formed from the photon beam by a 3 mm Converter is E ≈ 40 ± 9 MeV. If we take into account the energy resolution of the beam of secondary electrons formed by the converter ($\sigma \approx 12\%$ [3]), the energy value is E ≈ 40 ± 8 MeV. This resolution is the method resolution and is determined by fluctuations in the magnitude of the average electron beam path at the thickness of the SD.

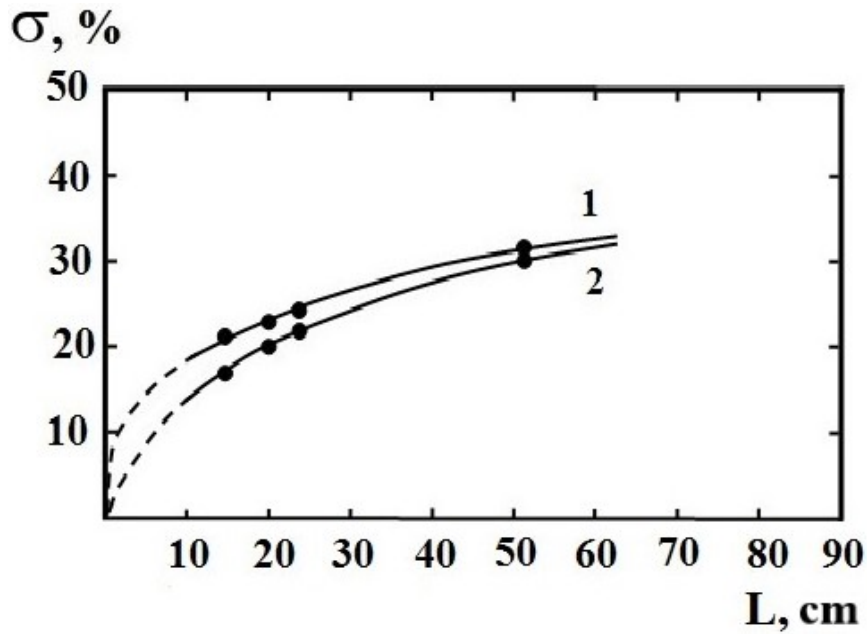

**Fig. 8** Dependence of the electron beam energy resolution measured by the "absorbed energy" method from the thickness L of scintillation detectors (1 - taking into account the influence of the copper converter; 2 - without the influence of the converter).

It should be noted that the type of amplitude spectra of SD varies depending on the electron beam energy. The Fig. 9 presents the amplitude spectra of SD with thickness of 20 cm (Fig.4) at electron beam energies up to the "inflection point" ($E_e$ = 9 MeV) (a), close to the "saturation point" ($E_e$ = 45 MeV) (b) and beyond the "inflection point" ($E_e$ = 145 MeV) (c). Clearly that at electron energies above the energy left by the particle at the thickness of the SD (Fig. 9c), the type of spectrum is actually determined by the Landau distribution [2].

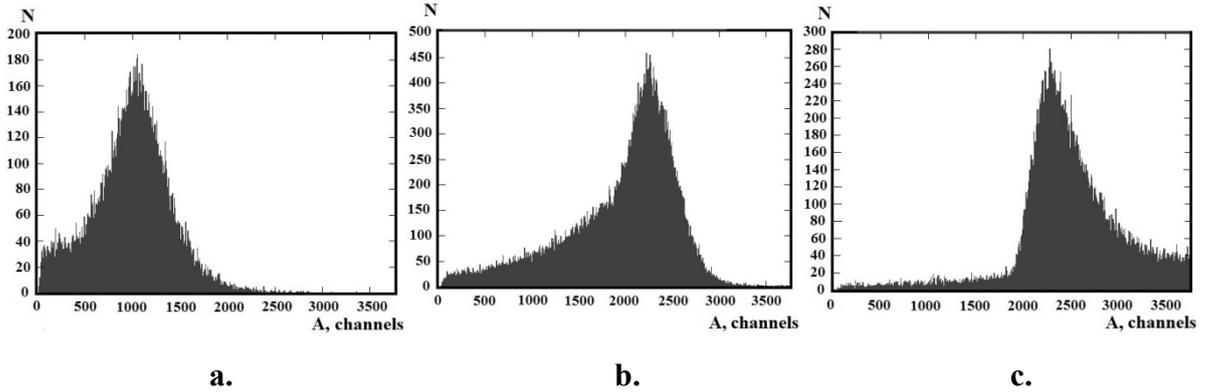

**a.**  **b.**  **c.**

**Fig. 9** Amplitude spectra of scintillation detector with thickness of 20 cm at electron beam energies up to the "inflection point" ($E_e$ = 9 MeV) (a), close to the "inflection point" ($E_e$ = 45 MeV) (b) and beyond the "inflection point" ($E_e$ = 145 MeV) (c).

The degree of spectrum change can be estimated using the coefficient method $\beta$ [5]. For each spectrum, the rate $\beta = \alpha_{right}/\alpha_{left}$ is calculated, where $\alpha_{right} = \sum_{i=m+1}^{m+n\sigma} N_i$ and $\alpha_{left} = \sum_{i=m-n\sigma}^{m} N_i$ is the number of events in the right and left parts of the spectrum, respectively, relative to the channel m, which determines the channel of the average amplitude value in the spectrum (σ is the error of the mean channel value (the full width of the amplitude spectrum at half its height); n is the number of standard deviations σ). One standard deviation n = 1 was used in the calculations. The dependence of the coefficient $\beta$ on the electron beam energy is presented in Fig. 10. It is seen that with increasing beam energy the spectrum begins to change, the maximum change in the spectrum is achieved at $E_e \approx 28$ MeV. The inflection point at which $\beta$ = 1 corresponds to $E_e \approx 40$ MeV. This means that at this point the electron tracks are optimally placed on the SD thickness. Otherwise, the tracks corresponding to the lower energy release will prevail over (left part of the spectrum, Fig. 9a) or tracks corresponding to high energy release (right side of the spectrum, Fig. 9b).

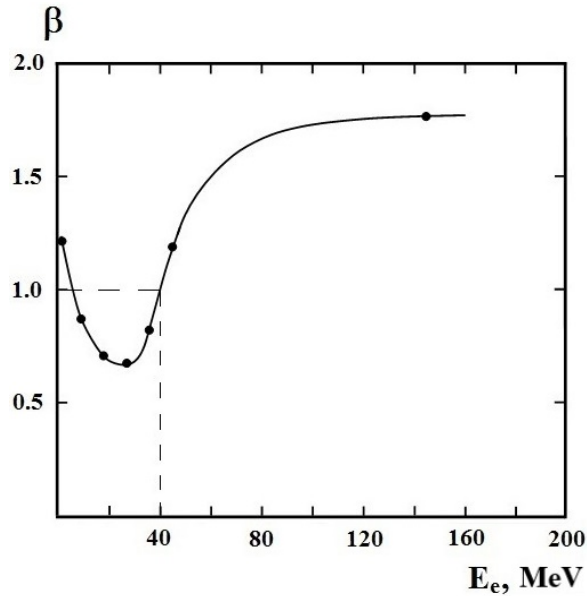

**Fig. 10** Dependence of the coefficient $\beta$ from the electron beam energy $E_e$.

The paper used polystyrene-based SD ($\rho \approx 1$ g / cm$^3$ [2]), the usage of SD from a denser material, for example, on the basis of crystals NaI(Tl), PbWO$_4$ and others ($\rho > 3$ g/cm$^3$ [2]) requires additional research. However, it can be assumed that to determine the electron beam energy of several hundred MeV, the SD thickness from a dense material (as well as from a light one) should be L < ~1-2$X_0$. At large thicknesses of the SD, electromagnetic showers begin to develop and energy is released from more than one electron track, which leads to an inaccurate determination of the beam energy by this method.

**Conclusion**

The presented method of the "absorbed energy" associated with the achievement of complete energy release of particles in a scintillation detector from a light material allows determining the energy of the electron beam and can be used in experimental activities. The area of use is preferred at energies of hundreds of MeV and the thickness of the scintillation detector is presumably up to ~100 cm (~2$X_0$), viz up to the region of the beginning of the electromagnetic shower development. In the energy range of tens of MeV and scintillation detector thicknesses up to ~20 cm (~0.5X0), the accuracy of determining the electron beam energy can be ~10 – 20%, which is close to the accuracy of determining the beam energy by traditional methods, such as the Cherenkov total absorption spectrometer [5].

This work was supported by grants from the Russian Foundation for Basic Research (NICA - RFBR) No. 18-02-40061 and No. 18-02-40079.